\documentclass[12pt]{article}
\pdfoutput=1
\usepackage{amsbsy}
\usepackage{amsfonts}
\usepackage{amsmath,amssymb}
\usepackage{bbold}
\usepackage{graphicx}

\newcommand{\sect}[1]{\setcounter{equation}{0}\section{#1}}

\newcommand{\eq}{\begin{equation}}
\newcommand{\eqa}{\begin{eqnarray}}  
\newcommand{\en}{\end{equation}}
\newcommand{\ena}{\end{eqnarray}}
\newcommand{\enn}{\nonumber \end{equation}}

\def\sk{\vskip .4cm}
\def\noi{\noindent}

\def\al{\alpha}
\def\be{\beta}
\def\ga{\gamma}

\let \part\partial

\def\part{\partial}

\def\sk{\vskip .4cm}

\def\noi{\noindent}

\def\X0{X^0}

\def\al{\alpha}
\def\ga{\gamma}

\def\square{{\,\lower0.9pt\vbox{\hrule \hbox{\vrule height 0.2 cm
\hskip 0.2 cm \vrule height 0.2 cm}\hrule}\,}}

\def\lb{\langle}
\def\rb{\rangle}

\def\Afat{\mathbb{A}}
\def\Bfat{\mathbb{B}}
\def\Pfat{\mathbb{P}}
\def\Qfat{\mathbb{Q}}

\def\Pcal{\cal {P}}

\def\noali{\al_i \mkern-15mu/}

\begin{document}

\begin{titlepage}

\vskip 2em
\begin{center}
{\Large \bf History entanglement entropy} \\[3em]

\vskip 0.5cm

{\bf
Leonardo Castellani}
\medskip

\vskip 0.5cm

{\sl Dipartimento di Scienze e Innovazione Tecnologica
\\Universit\`a del Piemonte Orientale, viale T. Michel 11, 15121 Alessandria, Italy\\ [.5em] INFN, Sezione di 
Torino, via P. Giuria 1, 10125 Torino, Italy\\ [.5em]
Arnold-Regge Center, via P. Giuria 1, 10125 Torino, Italy
}\\ [4em]
\end{center}

\begin{abstract}
\sk

A formalism is proposed to describe entangled quantum histories, and their entanglement entropy.  We define a history vector, living in a tensor space with basis elements corresponding to the allowed histories,
i.e. histories with nonvanishing amplitudes. The amplitudes are the components of the history
vector, and contain the dynamical information. Probabilities of measurement sequences, and resulting collapse, are given by generalized Born rules: they are all expressed by means of projections and scalar products involving the history vector.  Entangled history states are introduced, and a history density matrix is defined in terms of ensembles of history vectors. The corresponding history entropies (and history entanglement entropies for composite systems) are explicitly computed in two examples taken from
quantum computation circuits.

\end{abstract}

\vskip 6cm\
 \noi \hrule \vskip .2cm \noi {\small
leonardo.castellani@uniupo.it}

\end{titlepage}

\newpage
\setcounter{page}{1}

\tableofcontents

\sect{Introduction}

Formulations of quantum mechanics based on histories, rather than on states at a given time,
have their logical roots in the work of Feynman \cite{Feynman,FH} (see also the inspirational Chapter 32 of Dirac's book \cite{Dirac}), and could be seen as generalizations of the path-integral approach. A 
list with the references more relevant for the present work is given in \cite{histories1}-\cite{LC1}. 

We have seen in \cite{LC1} how to define a history operator on the Hilbert space ${\cal H}$ of a physical system, in terms of which to compute probabilities of successive measurements at times 
$t_1,...t_n$. In the present note we 
introduce a history vector, living in a tensor space ${\cal H} \odot {\cal H} ...\odot {\cal H}$,
where every ${\cal H}$ corresponds to a particular $t_i$. This vector contains
the same information as the history operator, but is more suited to define entanglement of histories,
and compute their density matrices and corresponding von Neumann entropies.

This approach is similar in spirit to the one advocated in ref.s \cite{histories11}-\cite{histories15}, but
with substantial differences. In \cite{histories11}-\cite{histories15} the scalar product between history states depends on chain operators containing information on evolution and measurements. In our framework
the algebraic structure does not depend on the dynamics, and {\it all} possible histories (not only ``consistent" sets) correspond to orthonormal vectors in ${\cal H} \odot {\cal H} ...\odot {\cal H}$. The dynamical information is instead encoded in the coefficients (amplitudes) multiplying the basis vectors.

The Born rules for probabilities and collapse are extended to history vectors in a
straightforward way. Every history vector has a pictorial representation in terms of
allowed histories, and its collapse after a measurement sequence entails the disappearance
of some histories. In this sense measurement ``alters the past", but never in a way to
endanger causality. As an illustration, the formalism is applied to
the entangler-disentangler and the teleportation quantum circuits.
\sk
The content of the paper is as follows. Chain operators and probabilities for multiple measurements at different times are recalled in Section 2. Section 3 introduces history amplitudes, an essential ingredient
in the definition of the history vector, given in Section 4. The generalized Born rules for probabilities of outcome sequences and collapse are derived, using
appropriate projectors on the history vector.
In Section 5 we propose a definition for history entanglement, based on a tensor product
between history states. Section 6 deals with density matrices, constructed using ensembles of
history vectors. This allows the computation of history entropy, and history entanglement entropy
for composite systems. Two examples based on
quantum computation circuits are provided in Section 7, and we calculate their history entanglement entropy. Section 8 contains some conclusions.

\sect{Chain operators and probabilities}

As recalled in \cite{LC1}, probabilities of obtaining sequences $\alpha= \alpha_1,\alpha_2,...\alpha_n$ of measurement results, starting from an initial state $|\psi\rb$, can all be expressed in terms of a chain operator $C_{\psi,\alpha}$. This operator encodes measurements at times $t_1,...t_n$ corresponding to projectors $P_{\alpha_{1}},...P_{\alpha_{n}} $, and unitary time evolution between measurements:
\eq
C_{\psi,\al} = P_{\alpha_{n}}  U(t_n,t_{n-1}) ~ P_{\alpha_{n-1}} ~ U(t_{n-1},t_{n-2})  \cdots P_{\alpha_{1}}~
U(t_1,t_0) P_{\psi} \label{chain}
\en
with $t_0 < t_1< \cdots < t_{n-1} < t_n$ and $P_{\psi}=|\psi\rb \lb \psi|$.  The $P_{\alpha_{i}} $ are projectors on eigensubspaces of 
observables, satisfying orthogonality and completeness relations:
\eq
P_{\alpha_{i}} P_{\beta_{i}}= \delta_{\alpha_{i},\beta_{i}} P_{\alpha_{i}},~~~I=\sum_{\alpha_{i}} P_{\alpha_{i}} \label{completeness}
\en
and $U(t_{i+1},t_i)$ is the evolution operator between times $t_i$ and $t_{i+1}$. 

The probability of obtaining the sequence $\alpha$ is given by
\eq
p(\psi, \al_1,\cdots \al_{n}) =  Tr (C_{\psi,\al} C_{\psi,\al}^\dagger ) \label{successive}
\en
and could be considered the ``probability of the history"  $ \psi,\al_1,\cdots \al_{n}$ . We can easily prove that the sum of all these probabilities gives 1:
\eq
\sum_\al  Tr ( C_{\psi,\al} C_{\psi,\al}^\dagger) =1
\en
by using the completeness relations in (\ref{completeness}) and unitarity of the $U(t_{i+1},t_{i}) $ operators.
We also find
\eq
\sum_{\al_{n}} p(\psi, \al_1,\al_2,\cdots  ,\al_{n})  =  p(\psi, \al_1, \al_2, \cdots,  \al_{n-1}) \label{sumrule0}
\en
However other standard sum rules for probabilities are not satisfied in general. For example relations of the type
\eq
\sum_{\al_2} p(\psi, \al_1,\al_2,\al_3)  =  p(\psi, \al_1, \al_3) \label{sumrule}
\en
hold only if the so-called {\it decoherence condition} is satisfied:
\eq
Tr (C_{\psi,\al} C_{\psi,\be}^\dagger) + c.c.= 0~~when~\al \not= \be  \label{decocondition}
\en
as can be checked on the example (\ref{sumrule}) written in terms of chain operators, and easily
generalized. Note that for chain operators the following is trivially true:
\eq
\sum_{\al_i} C_{\psi,\al_1,...\al_n} = C_{\psi,\al_1,...\noali,...\al_n} \label{reducedC}
\en
due to $\sum_{\al_i} P_{\al_i} = I$.

If all the histories we consider are such that the decoherence condition holds, they are said to form a {\it consistent} set, and can be assigned probabilities satisfying all the standard sum rules.

In general, histories do not form a consistent set: interference effects between them can be important, as in the case of the double slit experiment. For this reason we will not limit ourselves to consistent sets. Formula (\ref{successive}) for the probability of successive measurement outcomes holds true in any case.

\sect{Amplitudes}

\noi If $P_{\alpha_{n}}=|\al_n\rb\lb\al_n|$, i.e. the eigenvalue $\al_n$ is nondegenerate, the chain operator can be written as
\eq
C_{\psi,\al}=|\al_n\rb A(\psi,\al) \lb \psi |
\en
where
\eq
A(\psi,\al) = \lb \al_n| U(t_n,t_{n-1}) ~ P_{\alpha_{n-1}} ~ U(t_{n-1},t_{n-2})  \cdots P_{\alpha_{1}}~
U(t_1,t_0)  |\psi\rb \label{amplitude}
\en
is the {\sl amplitude} of the history $\psi,\al$, and 
\eq
|A(\psi,\al)|^2 = Tr(C_{\psi,\al} C^\dagger_{\psi,\al}) =p(\psi,\al)
\en
This easily generalizes to the case of a $g_n$-degenerate eigenvalue $\al_n$, with corresponding
(orthonormal) eigenvectors $|\al_n, i\rb$ ($i=1,...g_n$):
\eq
C_{\psi,\al}=\sum_i |\al_n,i\rb A_i(\psi,\al) \lb \psi |,~~~\sum_i |A_i (\psi,\al)|^2 = Tr(C_{\psi,\al} C^\dagger_{\psi,\al}) = p(\psi,\al)
\en
the amplitudes $A_i(\psi,\al)$ being given by formula  (\ref{amplitude}) where $\lb \al_n|$ is substituted by $\lb \al_n,i|$.
\sk
\noi A scalar product between chain operators can be defined as
\eq
(C_{\psi,\al},C_{\psi,\be})\equiv Tr (C_{\psi,\al} C_{\psi,\be}^\dagger)
\en
All the properties of a (complex) scalar product hold, in particular
\eq
(C_{\psi,\al},C_{\psi,\al}) = p(\psi,\al) = 0 ~~~\Longleftrightarrow ~~~C_{\psi,\al}=0
\en
{\bf Note :} if we divide the set $\al_1,...\al_{n-1}$ into two complementary sets
$\al_{i_1},...\al_{i_m}$ and $\al_{j_1},...\al_{j_p}$ with $m+p=n-1$, then
\eq
\sum_{\al_{j_1},...\al_{j_p}}  A(\psi,\al_1,...\al_{n-1},\al_n) = A(\psi,\al_{i_1},...\al_{i_m},\al_n) \label{reducedamp}
\en
because of the completeness relations in ({\ref{completeness}). This just rephrases property
(\ref{reducedC}) for chain operators, with the difference that $\al_n$ is never summed on since
it enters the amplitude (\ref{amplitude}) as a bra rather than as a projector. 

\sect{History vector, probabilities and collapse}

Consider a physical system in the state $|\psi\rb$ at time $t_0$ and devices that can be activated at times $t_1,...t_n$ to measure
given observables, with projectors on eigensubspaces as in (\ref{completeness}). Before any measurement, the system can be described by a {\sl history vector}, living in $n$-tensor space 
\eq
|\Psi\rb = \sum_\al A(\psi,\al) |\al_1\rb \odot ... \odot |\al_n\rb  \label{historyvector}
\en
where the coefficients $A(\psi,\al)$ are given by the amplitudes of the histories $\al = \al_1,...\al_n$,
computed as in the previous Section, and $|\al_k\rb$ are a basis of orthonormal vectors at each time $t_k$. 
If no degeneracy was present, these vectors would be just the eigenvectors of the observable(s) measured at time $t_k$. If the $\al_k$ ($k<n$) eigenvalues are degenerate, the information on degeneracy is lost in the symbol $|\al_k\rb$, but is contained in the amplitude $A(\psi,\al)$, where the projectors $P_{\al_k}$ on the eigensubspaces are present.  In case $\al_n$ is degenerate, the sum on $\al$ in (\ref{historyvector}) must
include the degeneracy index $i$, and (\ref{historyvector}) will be short for
\eq
|\Psi\rb = \sum_{\al,i} A_i(\psi,\al) |\al_1\rb \odot ... \odot |\al_{n-1}\rb \odot |\al_n,i\rb  \label{historyvectordeg}
\en
{\bf Note :}  In the following we will assume for simplicity that $\al_n$ is nondegenerate: all the results 
generalize easily to the degenerate case, usually by summing on the index $i$.
\sk
\noi The ``time product" $\odot$ has all the properties of a tensor product.
The symbol $\otimes$ (or just a blank) will be reserved for tensor products between states of subsystems at the same time $t_k$.
The vector is normalized since
\eq
\lb \Psi | \Psi \rb= \sum_\al |A(\psi,\al)|^2 = 1
\en
The {\it history content} of the system is defined to be the set of histories $\al = \al_1,...\al_n$ contained in
$|\Psi\rb$, i.e. all histories having nonvanishing amplitudes.
\sk
Probabilities of measuring sequences $\al=\al_1,...\al_n$ are given by the familiar formula
\eq
p(\psi,\al)= \lb \Psi | \mathbb{P}_\al |\Psi\rb = |A(\psi,\al)|^2.   \label{probvector}
\en
with 
\eq
\mathbb{P}_\al = |\al_1\rb\lb\al_1| \odot ... \odot  |\al_n\rb\lb\al_n| 
\en
Formula (\ref{probvector}) holds for sequences of measurements occurring at {\sl all} times $t_1,...t_n$. 
\sk
What is the effect of a sequence of measurements with results $\al_1,...\al_n$ on the system described by $|\Psi\rb$ ? We can characterize this effect as a {\sl collapse} of the history vector, implemented
mathematically by $\mathbb{P}_\al$. This projection collapses
the state $|\Psi\rb$ into the basis vector $ |\al_1\rb \odot ... \odot |\al_n\rb $ up to a phase:
\eq
|\Psi\rb ~~~\longrightarrow ~~~{\mathbb{P}_\al |\Psi\rb \over \sqrt{\lb \Psi  |\mathbb{P}_\al |\Psi\rb}} =  |\al_1\rb \odot ... \odot |\al_n\rb 
\en
The basis vector describes a system that has been ``completely measured" with results $\al_1,...\al_n$.
 Another sequence of measurements of the same observables at times $t_i$ would yield the same results $\al_1,...\al_n$ with probability one, according to the rule (\ref{probvector}). 
 
A {\sl partial} measurement at times $t_{i_1},...t_{i_m}$ ($m < n$) yielding the sequence $\al_{i_1},...\al_{i_m}$
likewise projects the state vector $|\Psi\rb$ into 
\eq
|\Psi_\al \rb = {\mathbb{P}_\al |\Psi\rb \over \sqrt{\lb \Psi | \mathbb{P}_\al |\Psi \rb}} \label{partialmeasure}
\en
where now $\mathbb{P}_\al$
is the projector on the sequence  $\al_{i_1},...\al_{i_m}$, i.e. a tensor product of identity operators and projectors at times $t_{i_1},...t_{i_m}$:
\eq
\mathbb{P}_\al = I \odot...\odot |\al_{i_1}\rb\lb \al_{i_1} | \odot I \odot...\odot |\al_{i_m}\rb\lb \al_{i_m} |\odot I \odot... \label{Ppartial}
\en
Then  $|\Psi_\al \rb$ is given, up to a normalization,  by the expression (\ref{historyvector})  where the sum on $\al$ 
involves only the times $t_j$ different from $t_{i_1},...t_{i_m}$, the rest of the $\al$'s being fixed to the values $\al_{i_1},...\al_{i_m}$.
\sk
The projected history vector $|\Psi_\al\rb$ can be used to compute conditional probabilities. The probability of obtaining the results $\be_{j_1},...\be_{j_p}$
at times $t_{j_1},...t_{j_p}$, given that $\al_{i_1},...\al_{i_m}$ are obtained at times 
$t_{i_1},...t_{i_m}$ (with $j_1,...j_p$ and $i_1,...i_m$ having no intersection, and union coinciding with $1,...n$), is given by
\eq
p(\be|\al) = \lb \Psi_\al | \mathbb{P}_\be |\Psi_\al \rb
\en

Finally, to compute probabilities for sequences $\al_{i_1},...\al_{i_m}$ in partial measurements at times $t_{i_1},...t_{i_m}$, we need a ``shorter" history vector with a reduced number of factors in the $\odot$ product corresponding to the subset $t_{i_1},...t_{i_m}$.  This vector can be obtained from $\Pfat_{\al} |\Psi\rb$ (with $\Pfat_{\al}$ as in (\ref{Ppartial})) by using a further projection ${\cal P}$, defined on the basis vectors as:
\eqa
& & {\cal P}_{i_1,...i_m} |\al_1\rb \odot ...\odot |\al_n\rb  \equiv  |\al_{i_1} \rb \odot...\odot |\al_{i_m}\rb     
\label{timeprojector1}
\ena
 if $t_{i_1},...t_{i_m}$ contains $t_n$, and as
\eq
{\cal P}_{i_1,...i_m} |\al_1\rb \odot ...\odot |\al_n\rb  \equiv  |\al_{i_1} \rb \odot...\odot |\al_{i_m}\rb \odot |\al_n\rb
\label{timeprojector}
\en
if $t_{i_1},...t_{i_m}$ does not contain $t_n$. 
For example
\eqa
& & {\cal P}_{1,3,5} |\al_1\rb \odot  |\al_2\rb \odot  |\al_3\rb \odot  |\al_4\rb \odot  |\al_5\rb =  |\al_1\rb \odot  |\al_3\rb \odot |\al_5\rb \\
& & {\cal P}_{1,2} |\al_1\rb \odot  |\al_2\rb \odot  |\al_3\rb \odot  |\al_4\rb \odot  |\al_5\rb =  |\al_1\rb \odot  |\al_2\rb \odot |\al_5\rb
\ena
The action of $\Pcal$ is then extended by linearity on any $|\Psi\rb$. Applying it to 
the vector $\Pfat_{\al} |\Psi \rb$ yields, when $t_{i_1},...t_{i_m}$ contains $t_n$:
\eqa
& & {\cal P}_{i_1,...i_m} \Pfat_{\al} |\Psi \rb =  {\cal P}_{i_1,...i_m}   \sum_{\al_{j_1},...\al_{j_p}}  A(\psi,\al)
 |\al_1\rb \odot ... \odot |\al_n \rb  = \nonumber \\
 & & =  \left( \sum_{\al_{j_1},...\al_{j_p}}  A(\psi,\al) \right) |\al_{i_1} \rb \odot ... \odot |\al_{i_m} \rb  = 
A(\psi, \al_{i_1},...\al_{i_m}) ~|\al_{i_1} \rb \odot ... \odot |\al_{i_m} \rb \nonumber \\  \label{psiproj1}
\ena
where we have used eq. (\ref{reducedamp}) in the second line. Then the probability 
$ |A(\psi,  \al_{i_1},...\al_{i_m})|^2$ of obtaining the partial sequence $\al_{i_1},...\al_{i_m}$ can be 
expressed as a scalar product
\eq
p(\psi,  \al_{i_1},...\al_{i_m})= |A(\psi,  \al_{i_1},...\al_{i_m})|^2 =  \lb \Psi | \Pfat_\al  {\cal P}_{i_1,...i_m}^\dagger  {\cal P}_{i_1,...i_m} {\mathbb P}_{ \al} |\Psi\rb 
\en
where $ \lb\al_1| \odot ...\odot \lb \al_n| {\cal P}_{i_1,...i_m}^\dagger$ is the conjugate of (\ref{timeprojector1}) or (\ref{timeprojector}). Note that
\eq
\Qfat_{i_1,...i_m} \equiv {\cal P}_{i_1,...i_m}^\dagger {\cal P}_{i_1,...i_m}
\en
is a hermitian operator in $n$-tensor space, with matrix elements
\eq
\lb\al_1| \odot ...\odot \lb \al_n| \Qfat_{i_1,...i_m}  |\beta_1\rb \odot ...\odot |\beta_n\rb = \delta_{\al_{i_1} \beta_{i_1}} \cdots ~\delta_{\al_{i_m} \beta_{i_m}}
\en
When $t_{i_1},...t_{i_m}$ does not contain $t_n$, $\al_n$ must be contained in 
$\al_{j_1},...\al_{j_p}$, and we have
\eqa
& & {\cal P}_{i_1,...i_m} \Pfat_{\al} |\Psi \rb =  {\cal P}_{i_1,...i_m}   \sum_{\al_{j_1},...\al_{j_{p-1}},\al_n}  A(\psi,\al)
 |\al_1\rb \odot ... \odot |\al_n \rb  = \nonumber \\
 & & = \sum_{\al_n}  \left( \sum_{\al_{j_1},...\al_{j_{p-1}}}  A(\psi,\al) \right) |\al_{i_1} \rb \odot ... \odot |\al_{i_m} \rb \odot |\al_n\rb   = \nonumber \\
& & = \sum_{\al_n} A(\psi, \al_{i_1},...\al_{i_m},\al_n) ~|\al_{i_1} \rb \odot ... \odot |\al_{i_m} \rb \odot  |\al_n\rb  \nonumber \\  \label{psiproj2}
\ena
The sequence probability $ |A(\psi,  \al_{i_1},...\al_{i_m})|^2$ is given by the 
same scalar product:
\eq
 \lb \Psi | \Pfat_\al  {\Qfat}_{i_1,...i_m} {\mathbb P}_{ \al} |\Psi\rb = \sum_{\al_n} |A(\psi,\al_{i_1},...\al_{i_m},\al_n)|^2 =  |A(\psi,  \al_{i_1},...\al_{i_m})|^2
 \en
 in virtue of relation (\ref{sumrule0}). Therefore we have established the formula 
 \eq
p(\psi,  \al_{i_1},...\al_{i_m})= \lb \Psi | \Pfat_\al {\Qfat}_{i_1,...i_m} {\mathbb P}_{ \al} |\Psi\rb \label{probscalar}
\en
for any partial sequence $\al_{i_1},...\al_{i_m}$.
    \sk
 \noi {\bf Note:} when $m=0$ (and therefore $\al_n$ is contained in $\al_{j_1},...\al_{j_p}$), $\Pfat_\al =$
 identity in tensor space and ${\cal P}$ projects on $t_n$. The projected vector in (\ref{psiproj2}) 
 becomes $\sum_{\al_{n}} A(\psi, \al_{n}) |\al_{n}\rb$ and is just
 the (usual) state vector $|\psi(t_{n})\rb$ of the system at time $t_{n}$, since
 \eq
 |\psi(t_{n})\rb = U(t_{n},t_0) |\psi\rb =  \sum_{\al_{n}} |\al_{n}\rb \lb \al_{n}| U(t_{i_1},t_0) |\psi\rb =   \sum_{\al_{n}} A(\psi, \al_{n}) |\al_{n}\rb
 \en

 \noi In conclusion, {\sl probabilities for (sequences of) measurements at any times can be computed via scalar 
 products involving appropriate projections of the history vector} $|\Psi\rb$.

\sect{History entanglement}

It is useful to define a tensor product between history vectors of subsystems. On the basis history vectors
the product acts as
\eq
 (|\al_1\rb \odot ... \odot |\al_n\rb) ( |\be_1\rb \odot ... \odot |\be_n\rb) \equiv
  |\al_1\rb |\be_1 \rb \odot ... \odot |\al_n \rb | \be_n\rb
  \en
and is extended by bilinearity on all linear combinations of these vectors. No symbol is used 
for this tensor product, to distinguish it from the tensor product $\odot$ involving different times $t_k$.

This allows a definition of product history states, which are defined to be
expressible in the form:
\eq
 (\sum_\al A(\psi,\al) |\al_1\rb \odot ... \odot |\al_n\rb ) (  \sum_\be A(\psi,\be) |\be_1\rb \odot ... \odot |\be_n\rb ) \label{hproduct}
 \en
or, using bilinearity:
\eq
\sum_{\al,\be} A(\psi,\al) A(\psi,\be) |\al_1 \be_1 \rb \odot...\odot |\al_n \be_n \rb
\en
with $|\al_i \be_i \rb \equiv |\al_i \rb | \be_i \rb$ for short. A product history state is thus
characterized by factorized amplitudes $A(\psi,\al,\be) =  A(\psi,\al) A(\psi,\be)$.
\sk
If the history state cannot be expressed as a product, 
we define it to be {\sl history entangled}\footnote{This entanglement is quite different from the one considered in ref.s \cite{histories11}-\cite{histories14}, where it involves superpositions of history states (without need of a composite system), and should be
considered as a {\sl temporal} entanglement.}. In this case,
results of measurements on system A are correlated with those on system B and viceversa. Indeed
if the amplitudes $A(\psi,\al,\be)$ in the history state
\eq
|\Psi^{AB}\rb=\sum_{\al,\be} A(\psi,\al,\be) |\al_1 \be_1 \rb \odot...\odot |\al_n \be_n \rb
\label{historyAB}
\en
 are not factorized, the probability for Alice to
obtain the sequence $\al$ if Bob obtains the sequence $\be$ {\it depends} on $\be$, and viceversa,
this probability being proportional to $|A(\psi,\al,\be)|^2$. On the other hand, if the history state is a
product (\ref{hproduct}), the probability for Alice is $|A(\psi,\al)|^2$ and does not depend on $\be$
(and likewise for Bob). 

We have the following criterion for history entanglement: starting from an initial state $|\psi\rb$ at $t_0$,
we examine all intermediate states of the system at times $t_i$, given by repeated application
of the evolution operators $U(t_i,t_{i-1})$.  If at least one of these intermediate states is an entangled state, then the history state of the system is entangled. This is because an entangled state at time $t_i$ implies a correlation between measurements at time $t_i$, which would be impossible if 
the history amplitudes for Alice and Bob measurements were factorized. Note that an entangled initial state $|\psi\rb$ does not necessarily imply history entanglement, since $U(t_1,t_0)$ could disentangle it.
\sk
The history vector (\ref{historyAB}) describes a bipartite system where the observables being measured at times $t_i$ are local observables of the form $A_i \otimes I$, $I \otimes B_i$, with eigenvalues $\al_i$ and $\be_i$ respectively. This is not the most general history state
of a bipartite system: the observables can be chosen to be global operators $C_i$ acting on the whole AB, with eigenvalues $\ga_i$. Then the history state reads:
\eq
\Psi^{AB} = \sum_{\ga} A(\psi,\ga) |\ga_1 \rb \odot...\odot |\ga_n\rb
\en
In this case we cannot extract from $|\Psi^{AB}\rb$ individual histories for the subsystems A and B.
\sk
Finally, the correlations in an entangled history system can be distinguished from the ``classical" correlations due to a statistical ensemble of history states, as discussed in next Section.
\sk

\sect{Density matrix and history entropy}

A system in the pure history state $|\Psi\rb$ has the density matrix:
\eq
\rho = |\Psi\rb \lb \Psi |
\en
a positive operator satisfying $Tr(\rho) = 1$ (due to $\lb \Psi |\Psi \rb = 1$). A mixed history state has density matrix
\eq
\rho = \sum_i p_i  |\Psi_i\rb \lb \Psi_i |
\en
with $\sum_i p_i = 1$, and $\{ |\Psi_i \rb \}$ an ensemble of history states.

Probabilities of measuring sequences $\al = \al_1,...\al_n$ in history state $\rho$ are given by the standard formula:
\eq
p(\al_1,...\al_n) = Tr(\rho ~ \mathbb{P}_\al)
\en
(cf. equation (\ref{probvector}) for pure states).

A (partial) measurement as the one considered in eq. (\ref{partialmeasure}) 
projects the density matrix in the usual way:
\eq
\rho ~~~\longrightarrow~~~\rho_\al = |\Psi_\al \rb \lb \Psi_\al| ={ \mathbb{P}_\al ~\rho ~\mathbb{P}_\al 
\over Tr(\rho ~ \mathbb{P}_\al)}
\en
and the probability of obtaining the partial sequence $\al_{i_1},...\al_{i_m}$ is given by 
\eq
p(\al_{i_1},...\al_{i_m}) = Tr(\rho~\Pfat_\al {\Qfat}_{i_1,...i_m} \Pfat_\al)
\en
cf. formula (\ref{probscalar}).

If a measurement is performed on $|\Psi\rb = \sum_\al A(\psi,\al) |\al\rb$, but the result remains unknown, the density matrix becomes
\eq
\rho ~~~\longrightarrow~~~ \rho' = \sum_\al |A(\psi,\al)|^2 |\Psi_\al \rb \lb \Psi_\al|
 \en
 and describes a {\it mixed} history state.
 \sk
\noi  Consider now the following two history states:
 \sk
\noi1)  the pure history state:
 \eq
 |\Psi\rb=\sum_\al A(\psi,\al) |\al\rb  
 \en
 where $|\al\rb \equiv |\al_1\rb \odot ...\odot |\al_n\rb$. Its density matrix is 
 \eq
 \rho_{pure} = |\Psi\rb \lb \Psi | = \sum_{\al,{\al}'} A(\psi,\al) A(\psi,{\al}')^* |\al\rb \lb {\al}'|  \label{pure}
 \en      
\noi 2) the mixed history state
\eq
\rho_{mixed} =  \sum_\al |A(\psi,\al)|^2 |\al \rb \lb \al|  \label{mixed}
\en

\noi The probabilities of obtaining a sequence $\al$ are the same for the two states, so they cannot be distinguished by a sequence of measurements at times $t_1,...t_n$. Recall the similar situation
for ordinary state vectors, where for example the mixed state $\rho_{mixed} = {1\over2} |0\rb\lb 0| + {1\over2} |1\rb\lb 1|$ can be distinguished from the pure state $\rho_{pure}={1\over2} (|0\rb + |1\rb)(\lb 0|+ \lb 1|)$ by measurements in a basis different from the computational one. For history states
however we must stick to the given set of observables at each time $t_i$, which {\sl defines} the history vector. Indeed the measuring devices are {\sl part} of the history description of the quantum system. One can change description by changing the measuring apparati, but then one must compute the new amplitudes for the new histories. There is no straightforward operation on the old history state that relates it to the new one\footnote{Trying to express $|\al_i\rb$ in terms of eigenvectors of other observables in (\ref{historyvector}) leads to wrong history amplitudes, as one can easily verify in the case of a qubit with evolution $t_0 \rightarrow t_1 \rightarrow t_2$.}. 
\sk
But then, how can we distinguish between the two history states (\ref{pure}) and (\ref{mixed}) ? There is a way, by using {\sl partial} measurements. Indeed the probability of obtaining a given partial sequence $\al_{i_1},...\al_{i_m}$, given by formula (\ref{probscalar}), takes different values for the states (\ref{pure}) and (\ref{mixed}). In the pure state this probability reads:
\eq
p(\al_{i_1},...,\al_{i_m})= Tr (\rho_{pure} \Pfat_\al {\Qfat}_{i_1,...i_m} \Pfat_\al )= | A(\psi,\al_{i_1},...,\al_{i_m})|^2 =  \left| \sum_{\al_{j_1},...\al_{j_p}} A(\psi,\al) \right|^2
\en
whereas for the mixed state:
\eq
p(\al_{i_1},...,\al_{i_m})= Tr (\rho_{mixed} \Pfat_\al {\Qfat}_{i_1,...i_m} \Pfat_\al )= \sum_{\al_{j_1},...\al_{j_p}} |A(\psi,\al)|^2
\en
with $\al_{j_1},...\al_{j_p}$ complementary to $\al_{i_1},...,\al_{i_m}$. Thus the difference is due to 
sum of square moduli being different from square modulus of sum, and we can experimentally distinguish between $\rho_{pure}$ and $\rho_{mxed}$. 
\sk
\noi Consider now a system AB composed by two subsystems A and B, and devices measuring
observables $\Afat_i = A_i \otimes I$ and $\Bfat_i=I \otimes B_i$  at each $t_i$. Its history state is
\eq
|\Psi^{AB}\rb = \sum_{\al,\be} A(\psi,\al,\be) |\al_1 \be_1 \rb \odot...\odot |\al_n \be_n \rb
\label{PsiAB}
\en
where $\al_i,\be_i$ are the possible outcomes of a joint measurement at time $t_i$ of 
$\Afat$ and $\Bfat$.  The amplitudes $A(\psi,\al,\be)$ are computed using the general formula
(\ref{amplitude}), with projectors 
\eq
\Pfat_{\al_i,\be_i} = |\al_i,\be_i \rb\lb \al_i,\be_i|= |\al_i \rb\lb \al_i| \otimes |\be_i \rb\lb\be_i|
\en
corresponding to the eigenvalues $\al_i,\be_i$. The density matrix of AB is
\eqa
& & \rho^{AB} = |\Psi^{AB}\rb \lb \Psi^{AB}| = \nonumber \\
 & & ~~~~~ = \sum_{\al,\be,{\al}', {\be}'} 
A(\psi,\al,\be) A(\psi,{\al}',{\be}')^* |\al_1 \be_1 \rb \odot...\odot |\al_n \be_n \rb 
 \lb \al_1 \be_1 | \odot...\odot \lb \al_n \be_n | \nonumber\\
\label{rhoAB}
\ena
Applying here the discussion of the preceding paragraph, we see that if (\ref{PsiAB}) describes
an entangled state, the correlations between Alice $\al$ and Bob $\be$ sequences can be
distinguished from correlations due to a statistical ensemble. 
\sk
We can define {\sl reduced density matrices} by partially tracing on the subsystems:
\eq
\rho^A \equiv Tr_B  ( \rho^{AB} ), ~~~\rho^B \equiv Tr_A  ( \rho^{AB} ) \label{rhoA}
\en
In general $\rho^A$ and $\rho^B$ will not describe pure history states anymore. 

This definition makes sense only if the reduced density matrices can be used to compute statistics for
measurements on the subsystems. Note that the history vector describes joint measurements 
on A and B, and therefore the probability for Alice
to obtain a particular sequence $\al_1,...\al_n$ in measuring $A$ on her subsystem must be computed
taking into account that also B gets measured (the result being unknown to Alice). This probability
is therefore given by the sum
\eq
p(\alpha) = \sum_\beta p(\alpha,\beta) = \sum_\beta  |A(\psi,\al,\be) |^2 \label{psum}
\en
Let us check that we obtain the same answer using the reduced density operator for Alice. 
Taking the partial trace on B of (\ref{rhoAB}) yields:
\eq
\rho^A = \sum_{\alpha,{\alpha}',\beta} A(\psi,\alpha,\beta) A^* (\psi,{\alpha}',\beta) 
|\al_1  \rb \odot...\odot |\al_n \rb  \lb {\al}'_1| \odot...\odot \lb {\al}'_n  |,
\en
a positive operator with unit trace. The standard expression in terms of $\rho^A$ for Alice's probability to obtain the sequence  $\al$ is 
\eq
p(\alpha) = Tr(\rho^A \Pfat_\alpha)  \label{prho}
\en
with
\eq
\Pfat_\alpha = (P_{\al_1} \otimes I) \odot \cdots \odot (P_{\al_n} \otimes I),~~~P_{\al_i} =
 |\al_i\rb\lb\al_i| \label{Pfat}
\en
It is immediate to verify that indeed the probability as computed in (\ref{prho}) coincides with 
(\ref{psum}), and therefore the definition (\ref{rhoA}) gives the correct density matrices for the subsystems.
\sk
On the other hand,  the probability for Alice to obtain the sequence
$\al_1,...\al_n$  {\sl with no measurements on Bob's part} is different from (\ref{psum}).
Indeed, the history vector of the composite system is different, since only
Alice's measuring device is activated, and reads
\eq
|\Psi^{AB}\rb = \sum_\al A(\psi,\al) |\al_1  \rb \odot...\odot |\al_n \rb \label{PsiAB2}
\en
where the amplitudes $A(\psi,\al)$ are obtained from the general formula (\ref{amplitude})
using the projectors $P_{\al_i}$ of (\ref{Pfat}). Here the reduced density operator $\rho^A$ is simply
\eq
\rho^A =  \sum_{\al,{\al}'}  A(\psi,\al) A(\psi,{\al}')^*  |\al_1  \rb \odot...\odot |\al_n \rb 
 \lb \al_1  | \odot...\odot \lb \al_n |
 \en
 (the trace on B has no effect, since history vectors contain only results of Alice), and the probability of Alice finding the sequence $\al$ is
 \eq
 p(\al) = Tr(\rho^A \Pfat_\alpha) =  |A(\psi,\al)|^2
 \en
 differing in general from (\ref{psum}).
\sk
\noi {\bf Note:} the amplitudes in (\ref{PsiAB2}) can also be computed as $A(\psi,\al)=\sum_\be A(\psi,\al,\be)$ due to $\sum_{\be_i} |\be_i\rb\lb \be_i| = I \longrightarrow \sum_{\be_i} \Pfat_{\al_i,\be_i} = P_{\al_i} \otimes I$. Thus the difference of the two situations described above
is due to
\eq
 \sum_\beta  |A(\psi,\al,\be) |^2 \not=  |A(\psi,\al)|^2
 \en
in general. In particular cases the equality sign holds, for example when
the evolution operator is factorized $U=U^A \otimes U^B$, i.e. when A and B do not interact.
\sk
Finally, we can define the system {\sl history (von Neumann) entropy} as
\eq
S(\rho^{AB}) = - \rho^{AB} \log \rho^{AB}
 \en
and, when $\rho^{AB}$ is a pure history state, the {\sl history entanglement entropies} for subsystems A and B:
\eq
S(\rho^A) = - \rho^A \log \rho^A,~~S(\rho^B) = - \rho^B \log \rho^B
\en
All known properties of von Neumann entropy hold, since they depend on $\rho^{AB}$ being
a positive operator with unit trace, and $\rho^A$, $\rho^B$ reduced density operators 
obtained by partial tracing. Some of these properties will be verified in the examples of
next Section.

\sect{Examples}

In this Section we examine two examples of quantum systems evolving from a given initial state, and subjected to successive measurements.
They are taken from simple quantum computation circuits\footnote{A review on quantum computation can be found for ex. in \cite{QC} .} where unitary gates determine the evolution between measurements. 
Only two gates are used: the Hadamard one-qubit gate $H$ defined by:
\eq
H |0\rb = {1\over \sqrt{2}} (|0\rb + |1\rb),~~~H |1\rb = {1\over \sqrt{2}} (|0\rb - |1\rb)
\en
and the two-qubit $CNOT$ gate:
\eq
CNOT |00\rb = |00\rb,~CNOT |01\rb = |01\rb,~CNOT |10\rb = |11\rb,~CNOT |11\rb = |10\rb
\en
Quantum computing circuits in the consistent history formalism have been discussed for example 
in ref.s \cite{histories2,quantumcomputation1}.

\subsection{Entangler-disentangler}

\includegraphics[scale=0.35]{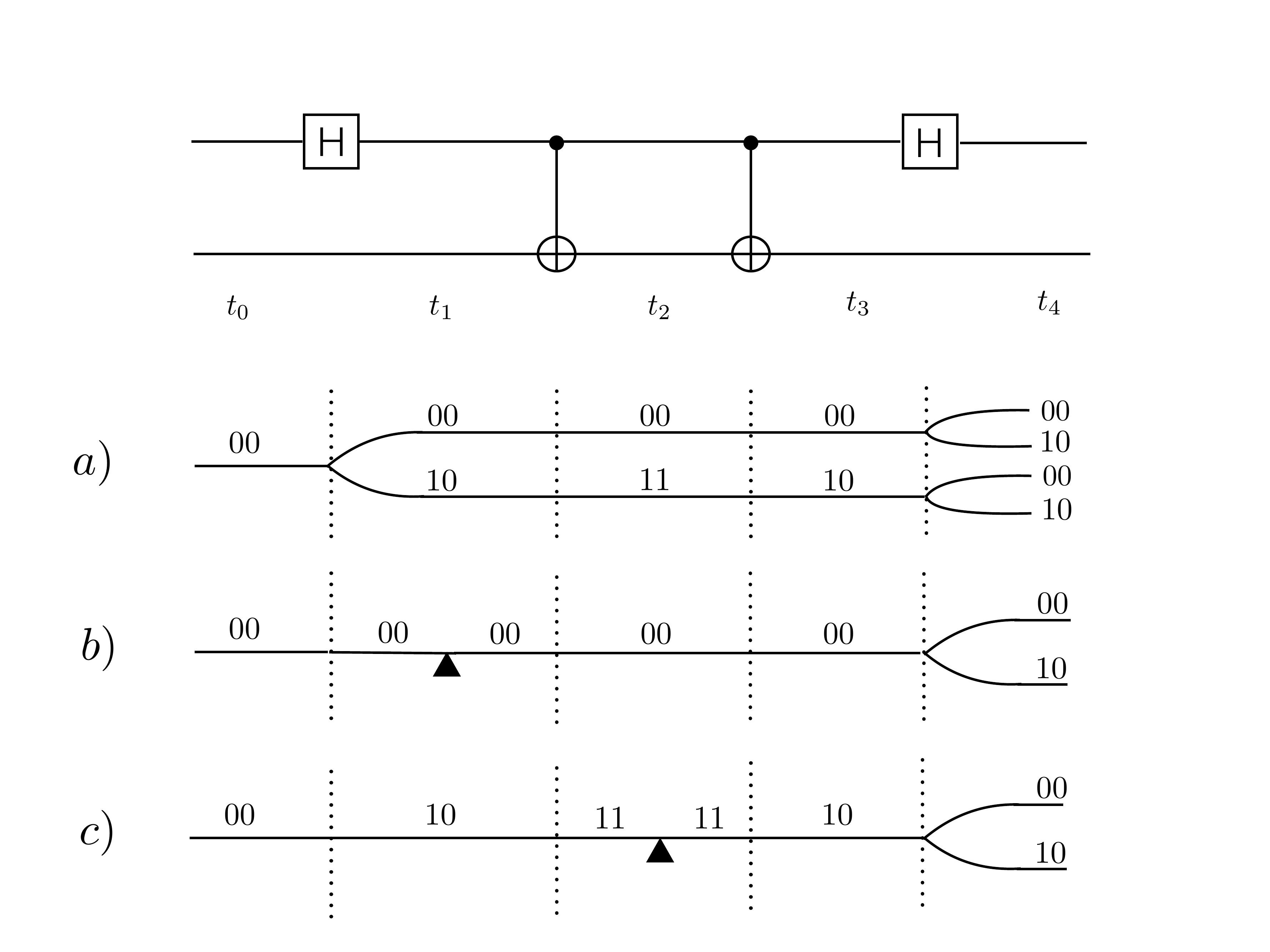}

\noi {\bf Fig. 1}  {\small The entangler - disentangler circuit, and some history diagrams for initial state $|00\rb$: a) no measurements, or Bob measures 0 at $t_1$;  b) Alice measures 0 at $t_1$; c) Alice measures 1 at $t_2$. Black triangles indicate measurements.}
\sk

If the initial state (at $t_0$) is $|00\rb$, the history state of the system before any measurements (at times $t_1,...t_4$) is given by
\eqa
& & |\Psi\rb = {1 \over 2} (|00\rb \odot |00\rb \odot  |00\rb \odot |00\rb +|00\rb \odot |00\rb \odot  |00\rb \odot |10\rb + \nonumber \\
& & ~~~~~~~~+|10\rb \odot |11\rb \odot  |10\rb \odot |00\rb -|10\rb \odot |11\rb \odot  |10\rb \odot |10\rb )
\ena
the amplitudes being given by formula (\ref{amplitude}), i.e.
\eqa
A(00,00,00,00,00) =\lb 00| (H \otimes I) | 00\rb\lb 00| CNOT | 00\rb\lb 00| CNOT | 00\rb\lb 00| (H \otimes I) | 00\rb= +{1 \over 2} \nonumber \\
A(00,00,00,00,10) = \lb 10| (H \otimes I) | 00\rb\lb 00| CNOT | 00\rb\lb 00| CNOT | 00\rb\lb 00| (H \otimes I) | 00\rb= +{1 \over 2} \nonumber \\
 A(00,10,11,10,00) =\lb 00| (H \otimes I) | 10\rb\lb 10| CNOT | 11\rb\lb 11| CNOT | 10\rb\lb 10| (H \otimes I) | 00\rb= +{1 \over 2}  \nonumber \\
 A(00,10,11,10,10) = \lb 10| (H \otimes I) | 10\rb\lb 10| CNOT | 11\rb\lb 11| CNOT | 10\rb\lb 10| (H \otimes I) | 00\rb= -{1 \over 2}  \nonumber\\
\ena
These amplitudes (or equivalently the history vector $|\Psi\rb$) encode all the necessary information to compute probabilities, according to the rules of Section 4. For example the probability of measuring any
of those four sequences is $1/4$, whereas the probability of measuring 10 at $t_4$ without measurements at $t_1,t_2,t_3$ is zero (the two histories with 10 at $t_4$ have opposite amplitudes and therefore interfere).

The history content of the system before measurements is displayed in diagram a) of Fig. 1. 
Measurements by Alice project the state $|\Psi\rb$ and reduce its history content as shown
in diagrams b) and c).

The unmeasured state $|\Psi\rb$ is history entangled, whereas the projected $|\Psi_\al\rb$ after
Alice measurements in diagrams b) and c) is a product history state.

The reduced density operator for Alice before measurements is
\eqa
& & \rho^A \equiv Tr_B  ( \rho^{AB} )= Tr_B  |\Psi \rb \lb\Psi| = \nonumber\\
& &~~ {1 \over 4} |0\rb \odot |0\rb \odot  |0\rb \odot |0\rb
\lb 0| \odot \lb 0| \odot  \lb 0| \odot \lb0|+ {1\over 4} |0\rb \odot |0\rb \odot  |0\rb \odot |1\rb
\lb 0| \odot \lb 0| \odot  \lb 0| \odot \lb1| \nonumber \\
& & + {1 \over 4} |1\rb \odot |1\rb \odot  |1\rb \odot |0\rb
\lb 1| \odot \lb 1| \odot  \lb 1| \odot \lb0|+ {1\over 4} |1\rb \odot |1\rb \odot  |1\rb \odot |1\rb
\lb 1| \odot \lb 1| \odot  \lb 1| \odot \lb1| \nonumber \\
& & + {1 \over 4} |0\rb \odot |0\rb \odot  |0\rb \odot |0\rb
\lb 0| \odot \lb 0| \odot  \lb 0| \odot \lb1|+ {1\over 4} |0\rb \odot |0\rb \odot  |0\rb \odot |1\rb
\lb 0| \odot \lb 0| \odot  \lb 0| \odot \lb0| \nonumber \\
& & -{1 \over 4} |1\rb \odot |1\rb \odot  |1\rb \odot |0\rb
\lb 1| \odot \lb 1| \odot  \lb 1| \odot \lb1|-  {1\over 4} |1\rb \odot |1\rb \odot  |1\rb \odot |1\rb
\lb 1| \odot \lb 1| \odot  \lb 1| \odot \lb0| \nonumber \\
\ena
or, in simplified notations:
\eq
 \rho^A = {1\over 2} {|0000\rb + |0001\rb \over \sqrt{2}}~ {\lb 0000| + \lb 0001| \over \sqrt{2}}  + {1\over 2} {|1111\rb -  |1110\rb \over \sqrt{2}}~ {\lb 1111| - \lb 1110| \over \sqrt{2}}
 \en
where $|0000\rb \equiv  |0\rb \odot |0\rb \odot  |0\rb \odot |0\rb$ etc. This density operator
describes a mixed history state, with an ensemble of two history vectors 
\eq
|000+\rb =  {|0000\rb + |0001\rb \over \sqrt{2}},~~~|111-\rb \equiv {|1111\rb -  |1110\rb \over \sqrt{2}}
\en
with equal probabilities $=1/2$. The reduced density matrix can be used to compute statistics for
Alice measurements. The AB system entropy is zero, since it is in a pure state, but 
the entropy corresponding to $\rho^A$ (the entropy ``seen" by Alice) is
\eq
S(\rho^A) = -Tr (\rho^A \log \rho^A) = - 2 ({1 \over 2} \log {1 \over 2}) = 1
\en
since $\rho^A$ has two nonzero eigenvalues equal to ${1\over 2}$. This is consistent 
with $\rho^A$ describing a mixed history state.

The reduced density operator for Bob is easily computed:
\eq
\rho^B = Tr_A (\rho^{AB}) = {1\over 2} |0000\rb \lb 0000| + {1 \over 2} |0100\rb \lb 0100\rb
\en
describing a statistical ensemble of the two histories $|0000\rb$ and $|0100\rb$ with equal probabilities $=1/2$, and history entropy $S(\rho^B)=S(\rho^A)=1$.

Note that without measurements the circuit is simply the identity circuit for two qubits, so the initial state 00 can only propagate to 00 at time $t_4$. The situation is different when intermediate measurements
are performed, as depicted in diagrams b) and c). In these cases also the state 10 at time $t_4$ becomes available.

\subsection{Teleportation}

The teleportation circuit \cite{teleportation}  is the three-qubit circuit given in Fig. 3, where the upper two qubits belong to Alice, and the 
lower one to Bob.

\includegraphics[scale=0.45]{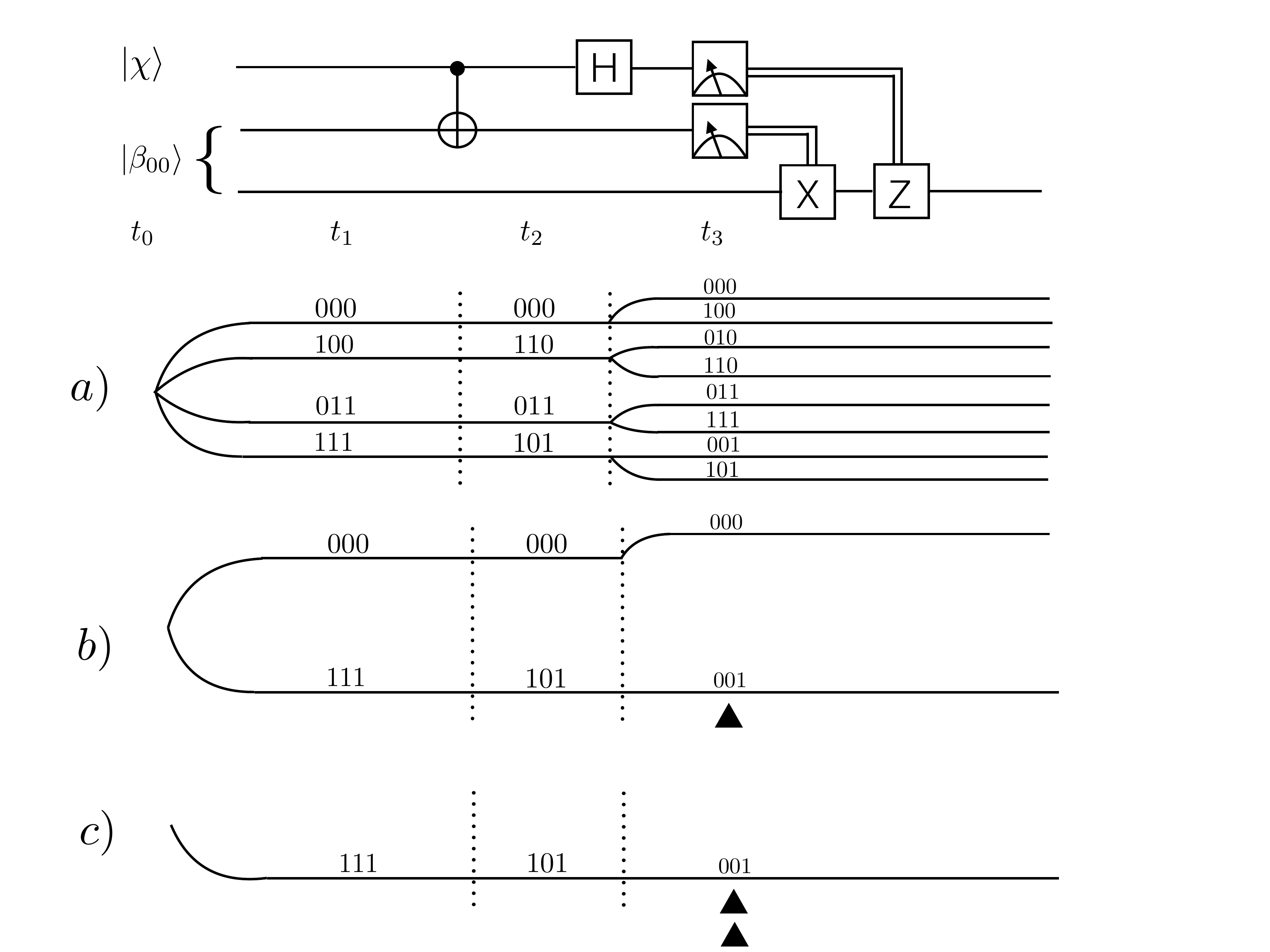}
\noi {\bf Fig. 2 } {\small Teleportation circuit: a) no measurements; b) Alice measures 00 at time $t_3$; c) at time $t_3$ Alice measures 00 
and Bob measures 1.}
\sk
\noi The initial state is a three-qubit state, tensor product of the single qubit $|\chi\rb = \al |0\rb + \be |1\rb$ to be teleported and the 2-qubit entangled Bell state
$|\be_{00}\rb = {1 \over \sqrt{2}} (|00 \rb + |11 \rb$. Before any measurement, the history vector contains
8 histories:
\eqa
& & |\Psi\rb = {1 \over 2} (\al |000\rb \odot |000\rb \odot  |000\rb  - \al|000\rb \odot |000\rb \odot  |100\rb  + \nonumber \\
& & ~~~~~~~~+ \be |100\rb \odot |110\rb \odot  |010\rb - \be |100\rb \odot |110\rb \odot  |110\rb \nonumber \\
& & ~~~~~~~~+\al |011\rb \odot |011\rb \odot  |011\rb - \al |011\rb \odot |011\rb \odot  |111\rb \nonumber \\
& & ~~~~~~~~+\be |111\rb \odot |101\rb \odot  |001\rb -\be |111\rb \odot |101\rb \odot  |101\rb \nonumber \\
\label{historytel}
\ena
the amplitudes being given by 
\eq
A(\chi \otimes \be_{00},\al_1,\al_2,\al_3) = \lb \al_3 | H_1 P_{\al_2} {\rm CNOT}_{1,2} P_{\al_1} |\chi \otimes \be_{00} \rb 
\en
For example
\eq
A(\chi \otimes \be_{00},000,000,000) = \lb 000|  H_1  |000\rb\lb 000|  {\rm CNOT}_{1,2}  |000\rb\lb 000|  \chi \otimes \be_{00} \rb = \al / 2
\en
where $H_1 \equiv H \otimes I \otimes I$ and $ {\rm CNOT_{1,2}} \equiv {\rm CNOT} \otimes I$.  For the moment we do not take into account the $X$ and $Z$
gates, activated by the results of Alice measurements at $t_3$. The history vector has the representation given in Fig. 2a.

Suppose now that Alice measures her two qubits at time $t_3$, without any prior measurement. To
compute probabilities we need first to compute $\Pfat_{\al_3} |\Psi\rb$ where $\al_3$ can take the four values 00, 01, 10, 11. For example, if $\al_3 = 00$, then $\Pfat_{\al_3}= I \odot I \odot (P_{00} \otimes I)$, and
\eq
\Pfat_{\al_3=00} |\Psi\rb = {\al \over 2}  |000\rb \odot |000\rb \odot  |000\rb + {\be \over 2} |111\rb \odot |101\rb \odot  |001\rb
\en
Projecting on $t_3$ yields
\eq
{\cal P}_3  \Pfat_{\al_3=00} |\Psi \rb =  {\al \over 2}  |000\rb + {\be \over 2} |001\rb
\en
so that
\eq
p(\psi,\al_3 = 00)= \lb \Psi |   \Pfat_{\al_3=00} {\cal P}_3^\dagger {\cal P}_3  \Pfat_{\al_3=00} |\Psi \rb = {1 \over 4}(|\al|^2 + |\be|^2) = {1 \over 4}
\en 
The other three outcomes for $\al_3$ have the same probability = 1/4. 

Once Alice has obtained $00$ at $t_3$, corresponding to the projector $P_{\al_3} = P_{00} \otimes I$,
the history vector collapses into
\eq
|\Psi_\al \rb = {I \odot I \odot (P_{00} \otimes I) |\Psi\rb  \over \sqrt{\lb \Psi|I \odot I \odot (P_{00} \otimes I) |\Psi\rb }}
 =\al |000\rb \odot |000\rb \odot  |000\rb + \be |111\rb \odot |101\rb \odot  |001\rb \label{historytel2}
\en
and corresponds to the diagram b) in Fig. 2. With this vector we can compute the conditional probabilities
that Bob measures 0 or 1 at $t_3$, given that Alice has measured 00:
\eqa
& & p(0_B|00_A) = \lb \Psi_\al |I \odot I \odot I \otimes P_0 | \Psi_\al \rb = |\al|^2 \nonumber \\
& & p(1_B|00_A) = \lb \Psi_\al |I \odot I \odot I \otimes P_1 | \Psi_\al \rb = |\be|^2
\ena
To find the (usual) state vector of the system at time $t_3$ we project $|\Psi_\al\rb$ on $t_3$ with
the use of the ${\cal P}_3$ projector:
\eq
|\Psi'\rb = {\cal P}_3 |\Psi_\al\rb = \al |000\rb + \be |001\rb = |00\rb (\al |0\rb + \be |1\rb)
\en
and we see that  Bob's qubit is in the correctly teleported state $|\chi\rb= \al |0\rb + \be |1\rb$.
\sk

Similar arguments hold if Alice obtains $01$ or $10$ or $11$. In these cases Bob's qubit at time
$t_3$ is found to be in states that can be transformed into $|\chi\rb$ using $X$ and $Z$ gates, represented by the Pauli matrices $\sigma_x$ and $\sigma_z$ on the ($|0\rb$, $|1\rb$) basis.
\sk
Finally, if at time $t_3$ Alice measures 00 and Bob measures 1, the history vector $|\Psi\rb$ collapses into
\eq
|\Psi_\al\rb = {I \odot I \odot (P_{00} \otimes P_1) |\Psi\rb  \over \sqrt{\lb \Psi|I \odot I \odot (P_{00} \otimes P_1) |\Psi\rb }}
= |111\rb \odot |101\rb \odot  |001\rb. \label{historytel3}
\en
and corresponds to the diagram c) in Fig. 2.
\sk

The unmeasured history vector $|\Psi\rb$ in (\ref{historytel}) is entangled. The history vector $|\Psi_\al\rb$ in (\ref{historytel2}) after Alice measures 00 is likewise entangled, even if the (usual) state of the system at $t_3$ is a product state. Only the history state (\ref{historytel3}) is a product history state
$(|11\rb\odot|10\rb\odot|00\rb) \otimes (|1\rb\odot |1\rb\odot |1\rb)$.
\sk
\noi {\bf Density matrix and entropy}
\sk
The von Neumann entropy for the system before measurements is zero, since the system is in a pure history state.
The reduced history density matrix for Bob, before any measurement, is given in terms of the 
history vector $|\Psi\rb$ in (\ref{historytel}):
\eqa
 \rho^B = Tr_A (|\Psi\rb\lb\Psi|) &=&{1\over 2}(|0\rb\odot|0\rb\odot|0\rb \lb 0|\odot\lb 0|\odot \lb 0|+
|1\rb\odot|1\rb\odot|1\rb \lb 1|\odot\lb 1|\odot \lb 1|) \nonumber \\ \label{rhoB}
\ena
and does not depend on $\al$ and $\be$. It describes a mixed history state, with corresponding von Neumann entropy $S(\rho^B)=\log 2=1$. 

 If Alice measures her two qubits, without communicating her result, the density matrix of the system becomes
\eq
\rho^{AB} = \sum_\gamma |A(\psi,\gamma)|^2 |\gamma\rb \lb \gamma| \label{rhoafter}
\en
(the sum on $\gamma$ is over the 8 histories contained in the history vector $|\Psi\rb$) yielding a matrix with 4 eigenvalues equal to $|\al|^2/4$ and 4 eigenvalues equal to $|\be|^2/4$. Then the von Neumann entropy is
\eq
S(\rho^{AB})= -|\al|^2 \log {|\al|^2 \over 4} - |\be|^2 \log {|\be|^2 \over 4} =  -|\al|^2 \log {|\al|^2 } - |\be|^2 \log {|\be|^2}+2 
\en
Setting $p=|\al|^2$, the entropy $S(p)=2-p\log p-(1-p) \log (1-p)$ is maximum and equal to 
$\log 2+2=3$ when $p=1/2$, and is minimum and equal to $2$ when $p=0,1$.

The reduced density matrix for Bob computed from (\ref{rhoafter}) coincides with the one before measurements by Alice given in (\ref{rhoB}), as expected, since Alice's act of measuring cannot be detected by Bob (only the two qubits of Alice are interacting). The
corresponding von Neumann entropy is therefore the same: $S(\rho^B) = -\log (1/2) = 1$.

\sect{Conclusions}

History amplitudes, or equivalently chain operators, contain all the information necessary
to compute probabilities of outcome sequences when measuring a given physical system.
In the paper \cite{LC1} we proposed a pictorial way to represent the history content (i.e. the set
of all histories with nonvanishing amplitudes) encoded in a history operator, acting on the 
Hilbert space ${\cal H}$ of physical states. In the present paper amplitudes are used
to construct a history vector, living in a tensor product of multiple ${\cal H}$ copies, in terms of which all probabilities can be expressed via projections and scalar products. 

The formalism proposed here has two advantages with respect to the usual state vector description
of a physical system: 

1) it provides a convenient way to keep track of all possible histories of the system,
and of their reduction due to measurements. This can be translated into graphs that facilitate
intuition on how the system behaves under unitary time evolution and measurements at different times. 

2) it allows the definition of history entanglement, history entropy, and history entanglement entropy for
composite systems.

\section*{Acknowledgements}

We thank the referee for having prompted several improvements and clarifications. This work is supported by the research funds of the Eastern Piedmont University and INFN - Torino Section.

\vfill\eject
\end{document}